\documentclass[aps,prd,reprint,groupedaddress]{revtex4-1}

\usepackage{graphics,graphicx}
\usepackage{epstopdf}
\newcommand{\as}{\alpha_s}
\newcommand{\ul}[1]{\underline{#1}}
\usepackage{amsmath}
\def\eq#1{{Eq.~(\ref{#1})}}
\def\fig#1{{Fig.~\ref{#1}}}
\newcommand{\tr}{\mbox{tr}}

\newcommand{\llangle}{\Big\langle \!\! \Big\langle}
\newcommand{\rrangle}{\Big\rangle \!\! \Big\rangle}
\newcommand{\half}{{1\over 2}}

\begin{document}

\title{Helicity Evolution at Small $x$: Summary of Recent Developments}

\author{Yuri V. Kovchegov}
\affiliation{Department of Physics, The Ohio State
           University, Columbus, OH 43210, USA}
\email[Email: ]{kovchegov.1@osu.edu}
\author{Daniel Pitonyak}
\affiliation{Division of Science, Penn State University-Berks, Reading, PA 19610, USA}	
\email[Email: ]{dap67@psu.edu}
\author{Matthew D. Sievert}
\affiliation{Theoretical Division, Los Alamos National Laboratory, Los Alamos, NM 87545, USA}
\email[Email: ]{sievertmd@lanl.gov}

\date{\today}

\begin{abstract}
  We construct small-$x$ evolution equations which can be used to
  calculate quark and anti-quark helicity TMDs and PDFs, along with
  the $g_1$ structure function. These evolution equations resum powers
  of $\as \, \ln^2 (1/x)$ in the polarization-dependent evolution
  along with the powers of $\as \, \ln (1/x)$ in the unpolarized
  evolution which includes saturation effects. The equations are
  written in an operator form in terms of polarization-dependent
  Wilson line-like operators. While the equations do not close in
  general, they become closed and self-contained systems of non-linear
  equations in the large-$N_c$ and large-$N_c \, \& \, N_f$ limits. We
  construct a numerical solution of the helicity evolution equations
  in the large-$N_c$ limit. Employing the extracted intercept, we are
  able to predict directly from theory the behavior of the quark
  helicity PDFs at small $x$, which should have important
  phenomenological consequences. We also give an estimate of how much
  of the proton's spin may be at small $x$ and what impact this has on
  the so-called ``spin crisis.''
\end{abstract}

\maketitle

\section{Introduction}

These proceedings are based on the work done in
\cite{Kovchegov:2015pbl,Kovchegov:2016weo,Kovchegov:2016zex}.

Our aim is to derive perturbative QCD prediction for the asymptotic
small Bjorken $x$ behavior of the quark and gluon helicity
distribution functions and for related observables. In these
proceedings we will concentrate on the flavor-singlet quark helicity
distribution $\Delta q (x, Q^2)$. We will derive helicity evolution
equations resumming powers of $\as \, \ln^2 (1/x)$ with $\as$ the
strong coupling constant: this resummation is referred to as the
double-logarithmic approximation (DLA). These evolution equations
allow one to determine the leading perturbative behavior of the
small-$x$ asymptotics of $\Delta q (x, Q^2)$ (see
\cite{Kovchegov:2016weo,Kovchegov:2016zex}). Such theoretical input is
necessary to assist the efforts to determine the small-$x$ part of the
quark contribution to proton spin
\begin{align}
  \label{eq:net_spin}
  & S_q (Q^2) = \frac{1}{2} \, \int\limits_0^1 dx \, \Delta \Sigma (x,
  Q^2), \notag \\ & \Delta \Sigma (x, Q^2) = \left[ \Delta u + \Delta
    {\bar u} + \Delta d + \Delta {\bar d} + \ldots \right] \! (x,
  Q^2),
\end{align}
where the helicity parton distribution functions (hPDFs) are
\begin{align}
  \label{eq:hPDFs}
  \Delta f (x, Q^2) \equiv f^+ (x, Q^2) - f^- (x, Q^2).
\end{align}
The ultimate goal of determining the proton spin carried by quarks
[and the spin carried by the gluons, $S_G (Q^2)$] is to resolve the
proton spin crisis.


\section{The Observables}
\label{sec:observables}

\begin{figure*}[tbh]
\centering
\includegraphics[width= 0.85 \textwidth]{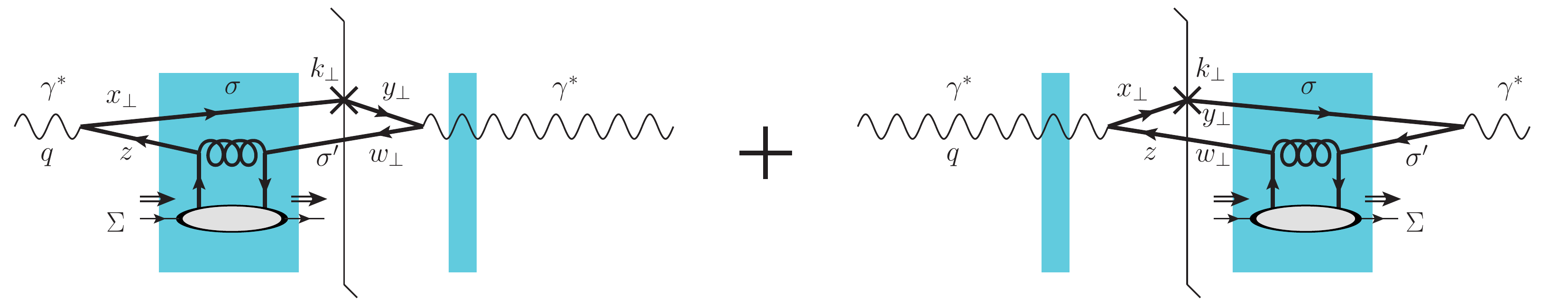}
\caption{Diagrams contributing to the small-$x$ SIDIS process on a
  longitudinally polarized target, and to quark helicity TMD $g^q_{1L}
  (x, k_T)$.}
\label{fig:helicity_TMD}
\end{figure*}

The small-$x$ helicity observables can be obtained by studying the
cross section for semi-inclusive deep inelastic scattering (SIDIS) on
a longitudinally polarized target, $\gamma^* + {\vec p} \to {\vec q} +
X$. The contributions are shown diagrammatically in
\fig{fig:helicity_TMD} (see \cite{Kovchegov:2015zha} for a
derivation).

The corresponding flavor-singlet quark helicity transverse
momentum-dependent parton distribution function (TMD) is
\cite{Kovchegov:2015pbl}
\begin{align}\label{eq:g1L}
  g_{1L}^S (x, k_T^2) = & \frac{8 \, N_c}{(2\pi)^6} \sum_f
  \int\limits_{z_i}^1 \frac{dz}{z} \int d^2 x_\perp \, d^2 y_\perp \,
  e^{-i \ul{k} \cdot (\ul{x} - \ul{y})} \notag \\ & \times \,
  \frac{\ul{x} - \ul{w}}{|\ul{x} - \ul{w}|^2} \cdot \frac{\ul{y} -
    \ul{w}}{|\ul{y} - \ul{w}|^2} \, d^2 w_\perp \, G_{\ul{x}, \ul{w}}
  (z) .
\end{align}
The notation is explained in \fig{fig:helicity_TMD}. Here $\ul{k} =
(k^x, k^y)$ denotes transverse vectors, with $k_\perp =
|\ul{k}|$. Variable $z$ denotes the fraction of the virtual photon's
longitudinal momentum carried by the anti-quark with $z_i =
\Lambda^2/s$, where $\Lambda$ is the infra-red (IR) cutoff, and $s$ is
the SIDIS center-of-mass energy squared. The object $G$ is the {\sl
  polarized dipole amplitude}, which is defined by
\cite{Kovchegov:2015pbl}
\begin{subequations}\label{e:Gdef}
  \begin{align} 
    \label{e:Gdef1} 
    & G_{10} (z) \equiv \frac{1}{2 N_c} \llangle \tr \left[ V_{\ul 0}
      V_{\ul 1}^{pol \, \dagger} \right] + \tr \left[V_{\ul 1}^{pol}
      V_{\ul 0}^\dagger \right] \rrangle (z), 
   %
   \\ \label{e:Gdef2}
   & G (x_{01}^2 , z) \equiv \int d^2 b \: G_{10} (z) ,
 \end{align}
\end{subequations}
where $\ul{b} = (1/2) (\ul{x}_1 + \ul{x}_0)$. The propagator of an
eikonal quark with polarization $\sigma$ in the background quark or
gluon field of the target proton is written as
\begin{align} 
  \label{e:Vsigma}
  V_{\ul x} (\sigma) \equiv V_{\ul x} + \sigma V_{\ul x}^{pol}
\end{align}
where 
\begin{align} 
  \label{e:Pexp}
  V_{\ul x} \equiv \mbox{P} \exp \left[ i g
    \int\limits_{-\infty}^\infty d x^+ \, A^- (x^+ , 0^-, \ul{x})
  \right] 
\end{align}
is the light-cone Wilson line, and $V^{pol}$ is the helicity-dependent
sub-eikonal correction. The double angle brackets indicate averaging
in the target wave function, with an inverse factor of center-of-mass
energy squared scaled out:
\begin{align}\label{redef0}
  \left\langle \ldots \right\rangle (z) = \frac{1}{z \, s} \,
  \left\langle \! \left\langle \ldots \right\rangle \!  \right\rangle
  (z).
\end{align}

The polarized dipole amplitude can be used to obtain other helicity
observables. The flavor-singlet quark helicity PDF, 
\begin{align}
  \Delta q^{S} (x, Q^2) \equiv \sum_f \left[ \Delta q^f (x, Q^2) +
    \Delta \bar{q}^f (x, Q^2) \right] ,
\end{align}
at small-$x$ is equal to
\begin{align}\label{eq:Deltaq}
  \Delta q^S (x, Q^2) = \frac{N_c}{2 \pi^3} \sum_f
  \int\limits_{z_i}^1 \frac{dz}{z} \int\limits_{\tfrac{1}{z
      s}}^{\tfrac{1}{z Q^2}} \frac{d |\ul{x} - \ul{w}|^2}{|\ul{x} -
    \ul{w}|^2} \, G(|\ul{x} - \ul{w}|^2 , z).
\end{align}
The $g_1$ structure function is
\begin{align}\label{eq:g1_all_twists}
  g^S_1 & (x, Q^2) = \frac{N_c}{2 \, \pi^2 \alpha_{EM}} \sum_f
  \int\limits_{z_i}^1\frac{dz }{z^2 (1-z)} \, \int d |\ul{x} -
  \ul{w}|^2 \notag \\ & \times \, \left[ \half \sum_{\lambda \sigma \sigma'} |
    \psi_{\lambda \sigma \sigma'}^T |^2_{(|\ul{x} - \ul{w}|^2 , z)} +
    \sum_{\sigma \sigma'} |\psi_{\sigma \sigma'}^L|^2_{(|\ul{x} -
      \ul{w}|^2 , z)} \right] \notag \\ & \times \, G(|\ul{x} -
  \ul{w}|^2 , z),
\end{align}
where $\psi^T$ and $\psi^L$ are the well-known light cone wave
functions for the $\gamma^* \to q \bar q$ splitting (see
e.g. \cite{Kovchegov:2015pbl}).

Our aim is to find the small-$x$ evolution equations for the polarized
dipole amplitude $G_{10} (z)$. Once $G_{10} (z)$ is found, we can use
Eqs.~\eqref{eq:g1L}, \eqref{eq:Deltaq} and \eqref{eq:g1_all_twists} to
construct the flavor-singlet quark helicity TMD, PDF and the $g_1$
structure function.


\section{Large-$N_c$ Limit}
\label{sec:large_nc}

Similar to the case of JIMWLK evolution and Balitsky hierarchy, the
general evolution equation for $G_{10} (z)$ does not close: on its
right-hand side it contains operator expectation values other than
$G_{10} (z)$. The operators on the right-hand side contain higher
number of Wilson lines than $G_{10} (z)$. This leads to the helicity
evolution analogue of the Balitsky hierarchy.

However, also similar to the unpolarized (BK) case, the evolution
equations become closed equations involving $G_{10} (z)$ in the
large-$N_c$ limit. In addition, specific to the helicity evolution
case at hand, evolution equations also close in the large-$N_c \, \&
\, N_f$ limit. Below we will first discuss the large-$N_c$ case.

Here we simply quote the results, referring the reader to the
derivation details in
\cite{Kovchegov:2015pbl,Kovchegov:2016zex}. Similar to
\cite{Itakura:2003jp}, our evolution equations also resum the
leading-logarithmic (LLA) powers of $\as \, \ln (1/x)$ by including
the BK/JIMWLK evolved unpolarized dipole $S$-matrix
\begin{align}\label{Sdef0}
  S_{01} (z) & = \frac{1}{2 N_c} \, \Big\langle \!\! \Big\langle
  \mbox{tr} \left[ V_{\ul{0}} \, V_{\ul{1}}^{\dagger} \right] +
  \mbox{tr} \left[ V_{\ul{1}} \, V_{\ul{0}}^{\dagger} \right]
  \Big\rangle \!\! \Big\rangle (z) \notag \\ & \approx \frac{1}{N_c}
  \, \Big\langle \!\!  \Big\langle \mbox{tr} \left[ V_{\ul{0}} \,
    V_{\ul{1}}^{\dagger} \right] \Big\rangle \!\!  \Big\rangle (z),
\end{align}
where we assume that 
\begin{align}\label{unp_tr}
  \mbox{tr} \left[ V_{\ul{0}} \, V^{\dagger}_{\ul{1}} \right] =
  \mbox{tr} \left[ V_{\ul{1}} \, V^{\dagger}_{\ul{0}} \right],
\end{align}
which is true at LLA. Note that LLA terms of the pure helicity
evolution are not systematically included in this approach: hence we
do not have a complete LLA calculation, and our results are strictly
correct only in the DLA limit where $S_{01} (z) =1$.

\begin{figure*}
\centering
\includegraphics[width= \textwidth]{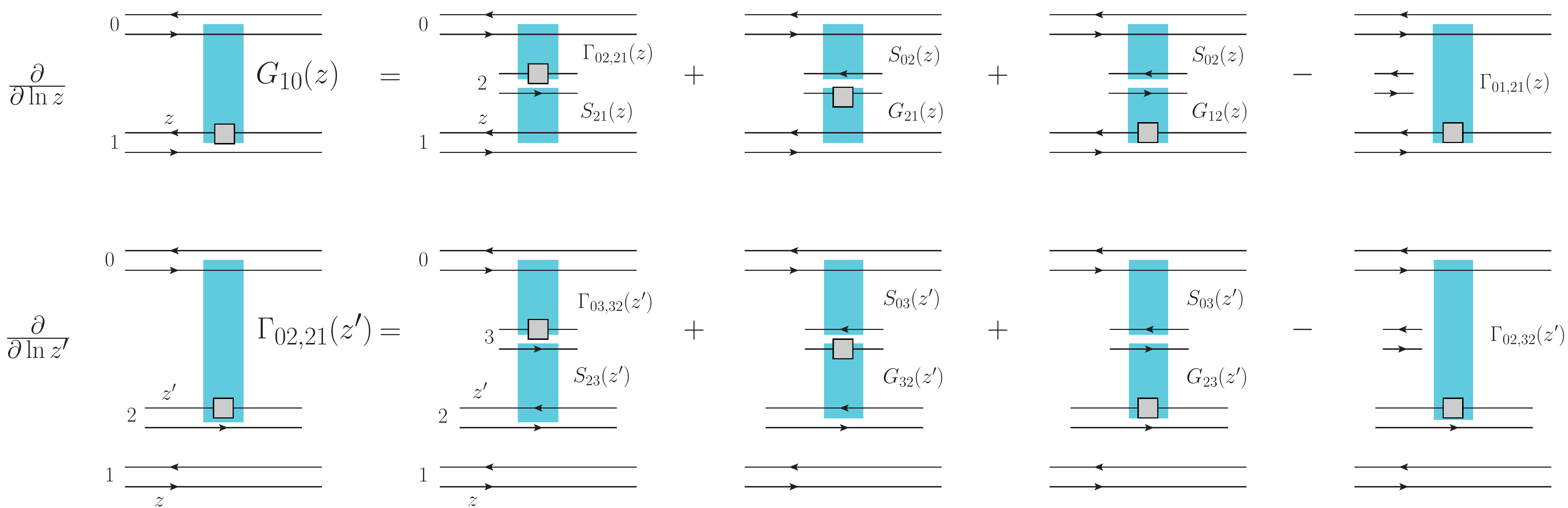}
\caption{Large-$N_c$ helicity evolution for the polarized dipole
  amplitude $G$ and the neighbor dipole amplitude $\Gamma$. For
  pictorial simplicity we do not show the contributions of the initial
  condition terms. Double lines denote gluons at large $N_c$. Only one
  of the virtual diagrams is shown (last diagram in each line):
  virtual corrections to the right of the shock wave are implied, but
  not shown explicitly. }
\label{Large-N_Evol}
\end{figure*}

The evolution equation for $G_{10} (z)$ is illustrated in the top line
of \fig{Large-N_Evol}. Note that the large-$N_c$ limit is
gluon-dominated: hence the dipole $10$ is made out of quark and
anti-quark lines of the large-$N_c$ gluon. The equation reads ($x_{ij}
= |\ul{x}_i - \ul{x}_j|$, $\rho'^2 = 1/(z' \, s)$)
\begin{align}\label{evol77}
  G_{10} (z) = & G_{10}^{(0)} (z)+ \frac{\as \, N_c}{2 \pi}
  \int\limits_{1/(s \, x_{10}^2)}^z \frac{d z'}{z'} \,
  \int\limits_{\rho'^2}^{x_{10}^2} \frac{d x^2_{21}}{x_{21}^2} \notag
  \\ \times & \, \left[ 2 \, \Gamma_{02, \, 21} (z') \, S_{21} (z') +
    2\, G_{21} (z') \, S_{02} (z') \right. \nonumber \\ & \left. + \,
    G_{12} (z') \, S_{02} (z') - \Gamma_{01, \, 21} (z') \right],
\end{align}
where $\Gamma_{02, \, 21} (z')$ is the new object (as compared to the
unpolarized evolution), characteristic of helicity
evolution. $\Gamma_{02, \, 21} (z')$ is the ``neighbor dipole''
amplitude. Its evolution is described in the bottom line of
\fig{Large-N_Evol}. As shown in the figure, the ``neighbor'' dipole
evolution continues in dipole $02$, but the information about the
dipole $21$ comes in through the transverse size integration
limit. (This is in contrast to unpolarized evolution, where the
evolution in, say, dipole $02$ does not depend on the size of the
dipole $21$ or on anything else outside the dipole $02$.) The
evolution for the neighbor dipole amplitude reads ($\rho''^2 = 1/(z''
\, s)$)
\begin{align}
  & \Gamma_{02, \, 21} (z') = \Gamma^{(0)}_{02, \, 21} (z') +
  \frac{\as \, N_c}{2 \pi} \int\limits_{1/(s \, x_{10}^2)}^{z'}
  \frac{d z''}{z''} \label{Gamma_evol} \\ & \times \,
  \int\limits_{{\rho''}^2}^{\mbox{min} \left\{ x_{10}^2 , x_{21}^2 \,
      z'/z'' \right\}} \frac{d x_{32}^2}{x_{32}^2} \, \left[ 2 \,
    \Gamma_{02, \, 32} (z'') \, S_{23} (z'') \right. \nonumber \\ &
  \left. + 2 \, G_{32} (z'') \, S_{03} (z'') + \, G_{23} (z'') \,
    S_{03} (z'') - \Gamma_{02, \, 32} (z'') \right]. \notag
\end{align}
Eqs.~\eqref{evol77} and \eqref{Gamma_evol}, when augmented by the BK
evolution for $S$, present a closed system of equations. The initial
conditions $G^{(0)}$ and $\Gamma^{(0)}$ are given by the Born-level
interactions, enhanced by multiple rescatterings which bring in
saturation effects.

In the strict DLA limit we can simplify Eqs.~\eqref{evol77} and
\eqref{Gamma_evol} by putting $S =1$ and assuming that $G_{21} =
G_{12}$. We obtain
\begin{subequations}\label{evol88}
\begin{align}
  G_{01} (z) & = G_{01}^{(0)} (z)+ \frac{\as \, N_c}{2 \pi}
  \int\limits_{1/(s \, x_{10}^2)}^z \frac{d z'}{z'} \\ & \times \,
  \int\limits_{{\rho'}^2}^{x_{10}^2} \frac{d x_{21}^2}{x_{21}^2} \,
  \left[  \Gamma_{02, \, 21} (z')  + 3 \, G_{21} (z')  \right], \notag \\
  \Gamma_{02, \, 21} (z') & = \Gamma^{(0)}_{02, \, 21} (z') +
  \frac{\as \, N_c}{2 \pi} \int\limits_{1/(s \, x_{10}^2)}^{z'}
  \frac{d z''}{z''} \label{Glin} \\ \times \, & \!\!\!\!\!
  \int\limits_{{\rho''}^2}^{\mbox{min} \left\{ x_{02}^2 , x_{21}^2 \,
      z'/z'' \right\}} \frac{d x_{32}^2}{x_{32}^2} \, \left[
    \Gamma_{02, \, 32} (z'') + 3 \, G_{23} (z'') \right]. \notag
\end{align}
\end{subequations}


\section{Large-$N_c \, \&  \, N_f$ Limit}
\label{sec:large_ncnf}

Helicity evolution equations also close in the large-$N_c \, \& \,
N_f$ limit. To write down these new evolution equations we need to
define a couple of new objects. In addition to $G_{10} (z)$ defined in
\eq{e:Gdef1} above, which is made out of quark and anti-quark lines of
gluons (with $x_1$ line polarized), let us define
\begin{align}\label{QGdef}
  A_{10} (z) = \frac{1}{2 N_c} \, \Big\langle \!\! \Big\langle
  \mbox{tr} \left[ V_{\ul{0}} \, V_{\ul{1}}^{pol \, \dagger} \right] +
  \mbox{tr} \left[ V^{pol}_{\ul{1}} \, V_{\ul{0}}^{\dagger} \right]
  \Big\rangle \!\! \Big\rangle (z)
\end{align}
with $x_1$ being a true quark or anti-quark polarized line and $x_0$
being the (anti-)quark line of the gluon, and
\begin{align}\label{Qdef}
  Q_{10} (z) = \frac{1}{2 N_c} \, \Big\langle \!\! \Big\langle
  \mbox{tr} \left[ V_{\ul{0}} \, V_{\ul{1}}^{pol \, \dagger} \right] +
  \mbox{tr} \left[ V^{pol}_{\ul{1}} \, V_{\ul{0}}^{\dagger} \right]
  \Big\rangle \!\! \Big\rangle (z)
\end{align}
with both $x_0$ and $x_1$ being true quark and anti-quark lines and
$x_1$ polarized. 

\begin{figure*}[htb]
\centering
\includegraphics[width= \textwidth]{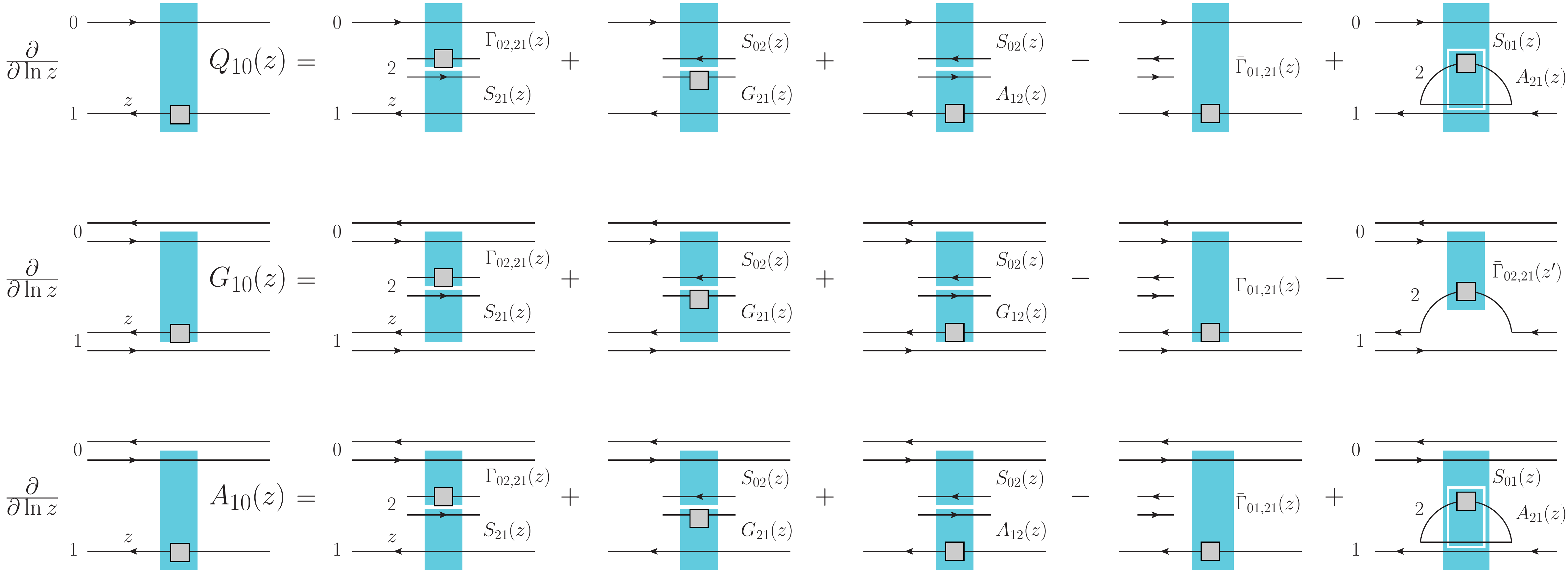}
\caption{Large-$N_c \, \& \, N_f$ helicity evolution for the polarized
  dipole amplitudes $Q$, $G$ and $A$.}
\label{Large-NcNf_Evol1}
\end{figure*}

Large-$N_c \, \& \, N_f$ evolution equations for $Q$, $G$ and $A$ are
illustrated diagrammatically in \fig{Large-NcNf_Evol1}, where again we
do not show the initial condition terms for simplicity. The equation
for $Q$ is
\begin{align}\label{Qevol}
  Q_{10} & (z) = Q_{10}^{(0)} (z) + \frac{\as \, N_c}{2 \pi}
  \int\limits_{z_i}^z \frac{d z'}{z'} \, \int\limits_{\rho'^2}^{x_{10}^2} \frac{d x^2_{21}}{x_{21}^2} \notag \\ & \times \left[ S_{21} (z') \, \Gamma_{02, \, 21} (z') + S_{02} (z') \, G_{21} (z')  \right. \notag \\ & \left. + S_{02} (z') \, A_{12} (z') - {\bar \Gamma}_{01, \, 21} (z') \right] \notag \\
  & + \frac{\as \, N_c}{4 \pi} \int\limits_{z_i}^z \frac{d z'}{z'} \,
  \int\limits_{\rho'^2}^{x_{10}^2 z/z'} \frac{d x^2_{21}}{x_{21}^2} \,
  S_{01} (z') \, A_{21} (z').
\end{align}
Equation \eqref{Qevol} is illustrated diagrammatically in the first
line of \fig{Large-NcNf_Evol1}. The equation for $G$ is now
\begin{align}\label{Gevol}
  G_{10} (z) & = G_{10}^{(0)} (z) + \frac{\as \, N_c}{2 \pi}
  \int\limits_{z_i}^z \frac{d z'}{z'} \,
  \int\limits_{\rho'^2}^{x_{10}^2} \frac{d x^2_{21}}{x_{21}^2} \notag
  \\ & \times \left[ 2\, S_{21} (z') \, \Gamma_{02, \, 21} (z') + 2\,
    S_{02} (z') \, G_{21} (z') \right. \notag \\ & \left. + S_{02}
    (z') \, G_{12} (z') - \Gamma_{01, \, 21} (z') \right] \notag \\ &
  - \frac{\as \, N_f}{4 \pi} \int\limits_{z_i}^z \frac{d z'}{z'} \,
  \int\limits_{\rho'^2}^{x_{10}^2 z/z'} \frac{d x^2_{21}}{x_{21}^2} \,
  {\bar \Gamma}_{02, \, 21} (z').
\end{align}
Note a new object, ${\bar \Gamma}_{02, \, 21}$, which is the neighbor
dipole amplitude with line $2$ being an actual polarized quark (or
anti-quark), and, unlike in $\Gamma_{02, \, 21}$, not a quark (or
anti-quark) line of a large-$N_c$ gluon. Equation \eqref{Gevol} is
illustrated diagrammatically in the second line of
\fig{Large-NcNf_Evol1}.

Finally, the evolution for $A_{01} (z)$ reads
\begin{align}\label{Aevol}
  & A_{10} (z) = A_{10}^{(0)} (z) + \frac{\as \, N_c}{2 \pi}
  \int\limits_{z_i}^z \frac{d z'}{z'} \,
  \int\limits_{\rho'^2}^{x_{10}^2} \frac{d x^2_{21}}{x_{21}^2} \notag
  \\ & \times \left[ S_{21} (z') \, \Gamma_{02, \, 21} (z') + S_{02}
    (z') \, G_{21} (z') \right. \notag \\ & \left. + S_{02} (z') \,
    A_{12} (z') - {\bar \Gamma}_{01, \, 21} (z') \right] \notag \\ &
  +\frac{\as \, N_c}{4 \pi} \int\limits_{z_i}^z \frac{d z'}{z'} \,
  \int\limits_{\rho'^2}^{x_{10}^2 z/z'} \frac{d x^2_{21}}{x_{21}^2} \,
  S_{01} (z') \, A_{21} (z').
\end{align}
It is depicted in the last line of \fig{Large-NcNf_Evol1}.

\begin{figure*}[bt]
\centering
\includegraphics[width= \textwidth]{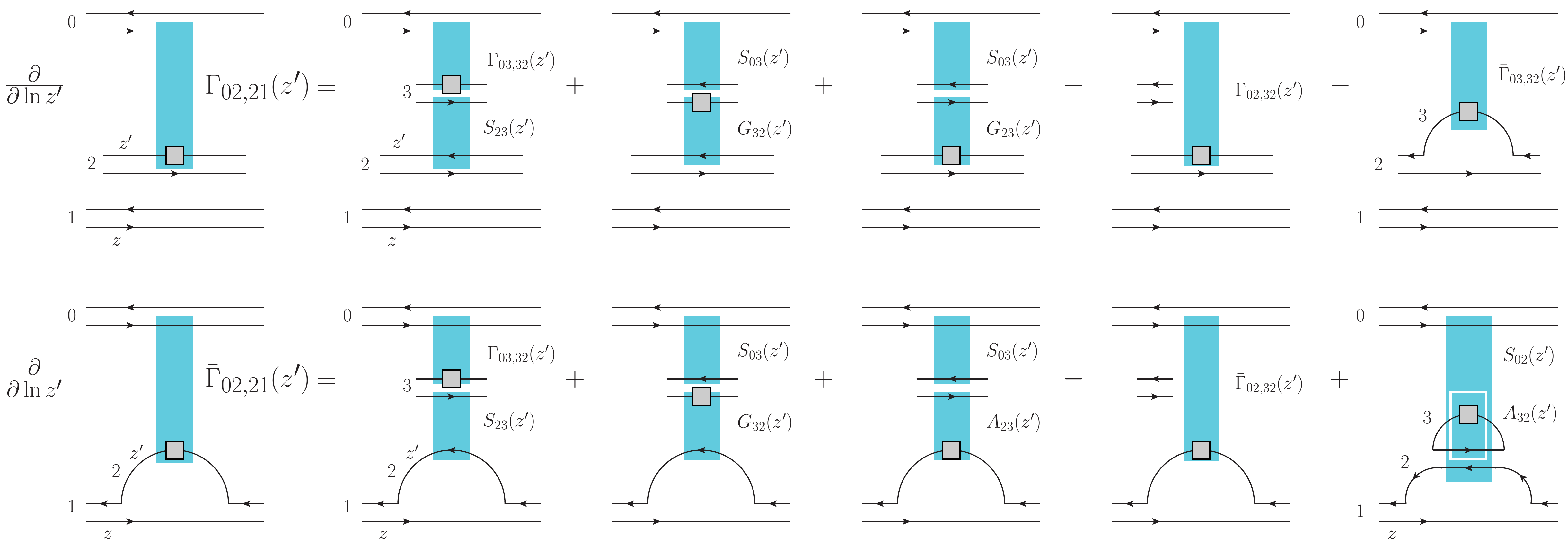}
\caption{Large-$N_c \, \& \, N_f$ helicity evolution for the polarized
  neighbor dipole amplitudes $\Gamma$ and $\bar \Gamma$.}
\label{Large-NcNf_Evol2}
\end{figure*}

Note that \eq{Gamma_evol} for the neighbor dipole amplitude also has
to be modified yielding
\begin{align}
  & \, \Gamma_{02, \, 21} (z') = \Gamma^{(0)}_{02, \, 21} (z') +
  \frac{\as \, N_c}{2 \pi} \int\limits_{z_i}^{z'} \frac{d z''}{z''}
  \!\!\!\!\!  \int\limits_{{\rho''}^2}^{\mbox{min} \left\{ x_{02}^2 ,
      x_{21}^2 \, z'/z'' \right\}} \!\!\!\!\!\!\!\!\!\!\!\!\!\!\!
  \frac{d x_{32}^2}{x_{32}^2} \notag \\ & \times \left[ 2\,
    \Gamma_{03, \, 32} (z'') \, S_{23} (z'') + 2 \, G_{32} (z'') \,
    S_{03} (z'') \right. \notag \\ & \left. + G_{23}
    (z'') \, S_{03} (z'') - \Gamma_{02, \, 32} (z'') \right] \notag \\
  & - \frac{\as \, N_f}{4 \pi} \int\limits_{z_i}^{z'} \frac{d
    z''}{z''} \, \int\limits_{\rho''^2}^{x_{21}^2 \, z'/z''} \frac{d
    x_{32}^2}{x_{32}^2} \, {\bar \Gamma}_{03, \, 32}
  (z'). \label{Gamma_evol2}
\end{align}
We also need an equation for $\bar \Gamma$:
\begin{align}
  & \, {\bar \Gamma}_{02, \, 21} (z') = {\bar \Gamma}^{(0)}_{02, \,
    21} (z') + \frac{\as \, N_c}{2 \pi} \int\limits_{z_i}^{z'} \frac{d
    z''}{z''} \!\!\!\!\! \int\limits_{{\rho''}^2}^{\mbox{min} \left\{
      x_{02}^2 , x_{21}^2 \, z'/z'' \right\}}
  \!\!\!\!\!\!\!\!\!\!\!\!\!\!\! \frac{d x_{32}^2}{x_{32}^2} \notag \\
  & \times \left[ \Gamma_{03, \, 32} (z'') \, S_{23} (z'') + G_{32}
    (z'') \, S_{03} (z'') \right. \notag \\ & \left. + A_{23} (z'') \,
    S_{03} (z'') - {\bar \Gamma}_{02, \, 32} (z'') \right] \notag \\ &
  + \frac{\as \, N_c}{4 \pi} \int\limits_{z_i}^{z'} \frac{d z''}{z''}
  \, \int\limits_{\rho''^2}^{x_{21}^2 \, z'/z'' } \frac{d
    x_{32}^2}{x_{32}^2} \, S_{02} (z') \, A_{32}
  (z'). \label{Gamma_evol3}
\end{align}
Both of these equations are diagrammatically illustrated in
\fig{Large-NcNf_Evol2}.

Equations~\eqref{Qevol}, \eqref{Gevol}, \eqref{Aevol},
\eqref{Gamma_evol2}, and \eqref{Gamma_evol3} are the large-$N_c \, \&
\, N_f$ helicity evolution equations which are DLA in
polarization-dependent terms, but also include LLA saturation
corrections through the $S$-matrices.

In the pure DLA limit we linearize all these equations by putting
$S=1$ in them (we again assume that $G_{01} = G_{10}$, which is true
for a large, longitudinally polarized target):
\begin{subequations}\label{Q_evol_lin}
\begin{align}\label{Qevol2}
  Q_{01} (z) = & \, Q_{01}^{(0)} (z) + \frac{\as \, N_c}{2 \pi}
  \int\limits_{z_i}^z \frac{d z'}{z'} \, \int\limits_{\rho'^2}^{x_{10}^2} \frac{d x^2_{21}}{x_{21}^2} \notag \\ & \times \left[ G_{12} (z') +  \Gamma_{02, \, 21} (z') + A_{21} (z') - {\bar \Gamma}_{01, \, 21} (z') \right] \notag \\
  & + \frac{\as \, N_c}{4 \pi^2} \int\limits_{z_i}^z \frac{d z'}{z'}
  \, \int\limits_{\rho'^2}^{x_{10}^2 z/z'} \frac{d x^2_{21}}{x_{21}^2} \, A_{21} (z'), \\
  G_{10} (z) = & \, G_{10}^{(0)} (z) + \frac{\as \, N_c}{2 \pi} \int\limits_{z_i}^z \frac{d z'}{z'} \, \int\limits_{\rho'^2}^{x_{10}^2} \frac{d x^2_{21}}{x_{21}^2} \notag \\ & \times \left[ \Gamma_{02, \, 21} (z') + 3 \, G_{12} (z') \right] \notag \\ & - \frac{\as \, N_f}{4 \pi} \int\limits_{z_i}^z \frac{d z'}{z'} \, \int\limits_{\rho'^2}^{x_{10}^2 z/z'} \frac{d x^2_{21}}{x_{21}^2} \, {\bar \Gamma}_{02, \, 21} (z'), \label{GGevol2} \\
  A_{01} (z) = & \, A_{01}^{(0)} (z) + \frac{\as \, N_c}{2 \pi}
  \int\limits_{z_i}^z \frac{d z'}{z'} \, \int\limits_{\rho'^2}^{x_{10}^2}
  \frac{d x^2_{21}}{x_{21}^2} \notag \\ & \times \left[ G_{12}
    (z') + \Gamma_{02, \, 21} (z') + A_{21} (z') - {\bar \Gamma}_{01,
      \, 21} (z') \right] \notag \\ & +\frac{\as \, N_c}{4 \pi}
  \int\limits_{z_i}^z \frac{d z'}{z'} \, \int\limits_{\rho'^2}^{x_{10}^2 z/z'}
  \frac{d x^2_{21}}{x_{21}^2} \, A_{12} (z').
\end{align}
\end{subequations}
The linearized equations for $\Gamma$ and $\bar \Gamma$ in the
large-$N_c \, \& \, N_f$ limit become
\begin{subequations}\label{Gam_evol_lin}
\begin{align}\label{Gamma_evol4}
  & \Gamma_{02, \, 21} (z') = \Gamma^{(0)}_{02, \, 21} (z') +
  \frac{\as \, N_c}{2 \pi} \int\limits_{z_i}^{z'} \frac{d z''}{z''} \\
  & \times \int\limits_{{\rho''}^2}^{\mbox{min} \left\{ x_{02}^2 , x_{21}^2 \, z'/z'' \right\}} \frac{d x_{32}^2}{x_{32}^2} \,  \left[ \Gamma_{03, \, 32} (z'')  + 3 \, G_{23} (z'') \right] \notag \\ & - \frac{\as \, N_f}{4 \pi} \int\limits_{z_i}^{z'} \frac{d z''}{z''} \, \int\limits_{\rho''^2}^{x_{21}^2 \, z'/z''} \frac{d x_{32}^2}{x_{32}^2} \,  {\bar \Gamma}_{03, \, 32} (z'), \notag \\
  & {\bar \Gamma}_{02, \, 21} (z') = {\bar \Gamma}^{(0)}_{02, \, 21}
  (z') + \frac{\as \, N_c}{2 \pi} \int\limits_{z_i}^{z'} \frac{d
    z''}{z''} \\ & \times \!\!\!\!\!\!\!
  \int\limits_{{\rho''}^2}^{\mbox{min} \left\{ x_{02}^2 , x_{21}^2 \,
      z'/z'' \right\}} \!\! \frac{d x_{32}^2}{x_{32}^2} \, \left[
    \Gamma_{03, \, 32} (z'') + G_{23} (z'') + A_{23} (z'')
  \right. \notag \\ & \left.- {\bar \Gamma}_{02, \, 32} (z'') \right]
  + \frac{\as \, N_c}{4 \pi} \int\limits_{z_i}^{z'} \frac{d z''}{z''}
  \, \int\limits_{\rho''^2}^{x_{21}^2 \, z'/z'' } \frac{d
    x_{32}^2}{x_{32}^2} \, A_{32} (z').
\end{align}
\end{subequations}
Note that in the large-$N_c \, \& \, N_f$ limit Eqs.~\eqref{eq:g1L},
\eqref{eq:Deltaq} and \eqref{eq:g1_all_twists} should contain $Q_{10}
(z)$ instead of $G_{10} (z)$.

Clearly in the large-$N_c$ / fixed-$N_f$ limit the linearized
equations for $G_{01} (z)$ and $\Gamma_{02, \, 21} (z')$ become a
closed system of equations \eqref{evol88} again, as employed in the
previous Subsection.  Since our final observable, quark helicity TMD
or hPDF, is related to $Q$, for the large-$N_c$ limit to be relevant,
$G$ should dominate (or at least be comparable to) $A$.

The linearized equations \eqref{Q_evol_lin} and \eqref{Gam_evol_lin},
when solved, should yield the helicity evolution intercept in the
large-$N_c \, \& \, N_f$ limit. Solution of Eqs.~\eqref{Q_evol_lin}
and \eqref{Gam_evol_lin} is left for the future (probably numerical)
work.


\section{Solution of the Large-$N_c$ Helicity Evolution Equations}

Let us now solve Eqs.~\eqref{evol88}. We start by defining new
coordinates,
\begin{align}
  &\hspace{1.3cm}\eta \equiv \ln \frac{z}{z_i} \, , \ \ \  \eta'\equiv \ln \frac{z'}{z_i} \,, \ \ \  \eta''\equiv \ln \frac{z''}{z_i} \, ,  \\[0.3cm]
  & s_{10} \equiv \ln \frac{1}{x_{10}^2 \Lambda^2}\,, \ \ \ s_{21}
  \equiv \ln \frac{1}{x_{21}^2 \Lambda^{2}}\,, \ \ \ s_{32} \equiv \ln
  \frac{1}{x_{32}^2 \Lambda^{2}}, \nonumber
\end{align}
as well as rescaling all $\eta$'s and $s_{ij}$'s, 
\begin{align}\label{resc}
  \eta \to \sqrt{\frac{2\pi}{\as N_c}} \ \eta\,, \ \ \ \ \ s_{ij} \to
  \sqrt{\frac{2\pi}{\as N_c}} \ s_{ij}\,.
\end{align}
Using these variables, we write the large-$N_c$ helicity evolution
equations \eqref{evol88} as
\begin{subequations}\label{GGeqns}
\begin{align} 
  & G (s_{10}, \eta) = G^{(0)} (s_{10}, \eta) + \int\limits_{s_{10}}^\eta d \eta' \int\limits_{s_{10}}^{\eta'} d s_{21} \label{e:G_redef} \\
  & \hspace{1.77cm} \times\,\left[ \Gamma (s_{10}, s_{21}, \eta') + 3 \, G (s_{21}, \eta') \right] \notag  \\
  & \Gamma (s_{10}, s_{21}, \eta') = \Gamma^{(0)} (s_{10}, s_{21}, \eta') + \int\limits_{s_{10}}^{\eta'} d \eta'' \label{e:Gam_redef} \\
  & \times\int\limits_{\mbox{max} \left\{ s_{10}, s_{21} + \eta'' -
      \eta' \right\}}^{\eta''} \hspace*{-1cm} d s_{32} \left[ \Gamma
    (s_{10}, s_{32}, \eta'') + 3 \, G (s_{32}, \eta'') \right]. \notag
\end{align}
\end{subequations}
Note that the ranges of the $s_{21}$ and $s_{32}$ integrations are
restricted to positive values of $s_{21}$ and $s_{32}$ as long as
$s_{10}$ is positive; therefore, we always stay above the IR cutoff
$\Lambda$ (in momentum space).  The initial conditions for
Eqs.~\eqref{GGeqns} are 
\cite{Kovchegov:2016zex}
\begin{align}
  G^{(0)}\! (s_{10}, \eta) &=  \Gamma^{(0)} \!(s_{10}, s_{21}, \eta) \nonumber\\
  &= \as^2\pi \tfrac{C_F}{N_c} \!\left[ C_F \, \eta -2 (\eta - s_{10})
  \right] ,\label{e:in_cond}
\end{align}
with $C_F = (N_c^2-1)/(2 N_c)$.  Since the equations at hand are
linear, and we are mainly interested in the high-energy intercept, we
can scale out $\as^2\pi\,C_F/N_c$.

In order to solve Eqs.~\eqref{GGeqns}, we first write down a
discretized version of them
\begin{subequations}\label{eqs_discr}
\begin{align}
  G_{ij} & \!=\! G^{(0)}_{ij} + \Delta \eta\,\Delta s \sum_{j'=i}^{j-1} \sum_{i' = i}^{j'}  \left[ \Gamma_{ii'j'} + 3 \, G_{i'j'} \right], \label{e:G_dis}\\
  \Gamma_{ikj} & \!=\! \Gamma^{(0)}_{ikj} + \Delta \eta\,\Delta s
  \!\sum_{j'=i}^{j-1} \ \sum_{i' = \mbox{max} \{ i, k+j'-j \}
  }^{j'}\!\! \left[ \Gamma_{ii'j'} + 3 \,
    G_{i'j'}\right],\nonumber\\ \label{e:Gam_dis}
\end{align}
\end{subequations}
where $G_{ij} \equiv G(s_i,\eta_j)$, $\Gamma_{ijk} \equiv \Gamma(s_i,
s_k, \eta_j)$, and
\begin{align}
\Delta \eta = \frac{\eta_{max}}{N_\eta}\,, \ \ \  \Delta s = \frac{\, s_{max}}{\, N_s} \,,
\end{align}
with $\eta_{max}$ the maximum $\eta$ value and $N_\eta$ the number of
grid steps in the $\eta$ direction, and likewise for $s_{max}$, $N_s$.
The discretized equations \eqref{eqs_discr} are exact in the limit
$\Delta\eta\,,\,\Delta s\to 0$ and $\eta_{max}\,,\,s_{max} \!\to
\infty$.  To optimize the numerics, we set $\eta_{max} = s_{max}$.

\begin{figure}[tb]
\centering \includegraphics[width=0.47 \textwidth]{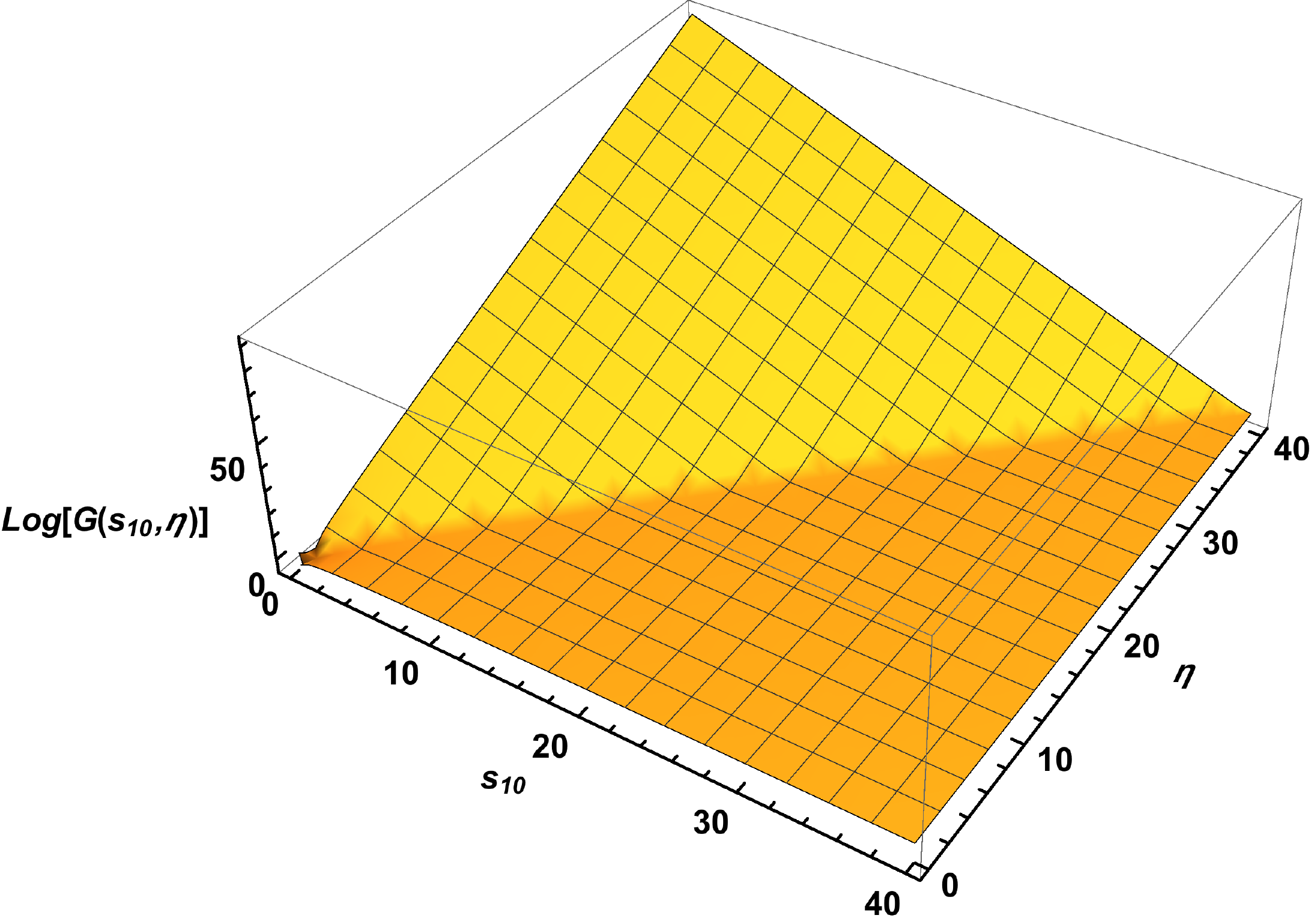}
\caption{The numerical solution of Eqs.~\eqref{GGeqns} for the
  polarized dipole amplitude $G$ plotted as a function of rescaled
  ``rapidity'' $\eta$ and transverse variable $s_{10}$.}
\label{f:G}
\end{figure}

With the discretized evolution equations \eqref{eqs_discr} in hand
(along with the initial conditions (\ref{e:in_cond}) suitably
discretized), we first choose values for $\eta_{max} = s_{max}$ and
$\Delta \eta = \Delta s$.  We then systematically go through the
$\eta$-$s$ grid in such a way that each $G_{ij}$ (and $\Gamma_{ijk}$)
only depends on $G,\Gamma$ values that have already been calculated.
Thus, we can determine $G_{ij}$ for each $i,j$. Our numerical solution
(for $\eta_{max} = 40$, $\Delta \eta = 0.05$) is plotted in \fig{f:G}.

We find
\begin{align} 
  \label{e:DeltaSigma}
  \Delta q^S(x,Q^2) \sim \Delta\Sigma(x, Q^2) \sim \left(\frac{1}
    {x}\right)^{\!\alpha_{h}}
\end{align}
with
\begin{align} 
  \label{e:alphah}
  \alpha_{h} = 2.31 \, \sqrt{\frac{\as N_c} {2\pi}}\, ,
\end{align}
where we have reinstated the factor $\sqrt{\as N_c/2\pi}$ originally
scaled out by Eq.~(\ref{resc}).  

We note that the value in Eq.~(\ref{e:alphah}) is in disagreement with
the ``pure glue'' intercept of $3.66\sqrt{\as N_c/2\pi}$ obtained by
BER \cite{Bartels:1996wc} by about $35\%$.  In
\cite{Kovchegov:2016zex} we identify DLA diagram contributions not
included by the authors of \cite{Bartels:1996wc} in their treatment of
the problem. We believe that omitting those diagrams limited the
resummation of \cite{Bartels:1996wc} to the leading-twist contribution
only. In comparison, our result \eqref{e:alphah} resums all twists at
small $x$. 

Interestingly, the leading twist approximation to $\alpha_P - 1$ in
BFKL evolution is larger than the exact all-twist intercept by about
$30\%$ \cite{KovchegovLevin}; it is possible that something similar is
occurring for helicity evolution.  In Ref.~\cite{Kovchegov:2016zex},
we have explored this possibility, performed various analytical
cross-checks of our helicity evolution equations, and compared to BER
where possible; we have not found any inconsistencies in our result.


\section{Impact on the proton spin}

In order to determine the quark and gluon spin based on
Eq.~(\ref{eq:net_spin}), one needs to extract the helicity PDFs.
There are several groups who have performed such analyses, e.g.,
DSSV~\cite{deFlorian:2009vb,deFlorian:2014yva},
JAM~\cite{Jimenez-Delgado:2013boa,Sato:2016tuz},
LSS~\cite{Leader:2005ci,Leader:2010rb,Leader:2014uua},
NNPDF~\cite{Ball:2013lla,Nocera:2014gqa}.  While the focus at small
$x$ has been on the behavior of $\Delta G(x,Q^2)$, there is actually
quite a bit of uncertainty in the size of $\Delta\Sigma(x,Q^2)$ in
that regime as well.

Let us define the truncated integral
\begin{align} 
  \label{e:run_int}
  \Delta\Sigma^{[x_{min}]}\!(Q^2)\equiv \int_{x_{min}}^1 \!dx
  \,\Delta\Sigma(x,Q^2)\,.
\end{align}
One finds for DSSV14~\cite{deFlorian:2014yva} that the central value
of the full integral $\Delta\Sigma^{[0]}\!(10\,{\rm GeV^2})$ is about
$40\%$ smaller than $\Delta\Sigma^{[0.001]}\!(10\,{\rm GeV^2})$.  The
NNPDF14~\cite{Nocera:2014gqa} helicity PDFs lead to a similar
decrease, although, due to the nature of neural network fits, the
uncertainty in this extrapolation is $100\%$.  On the other hand, for
JAM16~\cite{Sato:2016tuz} helicity PDFs the decrease from the
truncated to the full integral of $\Delta\Sigma(x,Q^2)$ seems to be at
most a few percent.  The origin of this uncertainty, and more
generally the behavior of $\Delta\Sigma(x,Q^2)$ at small $x$, is
mainly due to varying predictions for the size and shape of the sea
helicity PDFs, in particular $\Delta
s(x,Q^2)$~\cite{deFlorian:2009vb,deFlorian:2014yva,Jimenez-Delgado:2013boa,Sato:2016tuz,Ball:2013lla,Nocera:2014gqa,Aschenauer:2015ata}.
So far, the only constraint on $\Delta s(x,Q^2)$, and how it evolves
at small $x$, comes from the weak neutron and hyperon decay constants.
Therefore, there is a definite need for direct input from theory on
the small-$x$ intercept of $\Delta\Sigma(x,Q^2)$: this is what we have
provided in this Letter.

\begin{figure}[t]
\centering \includegraphics[scale=0.3]{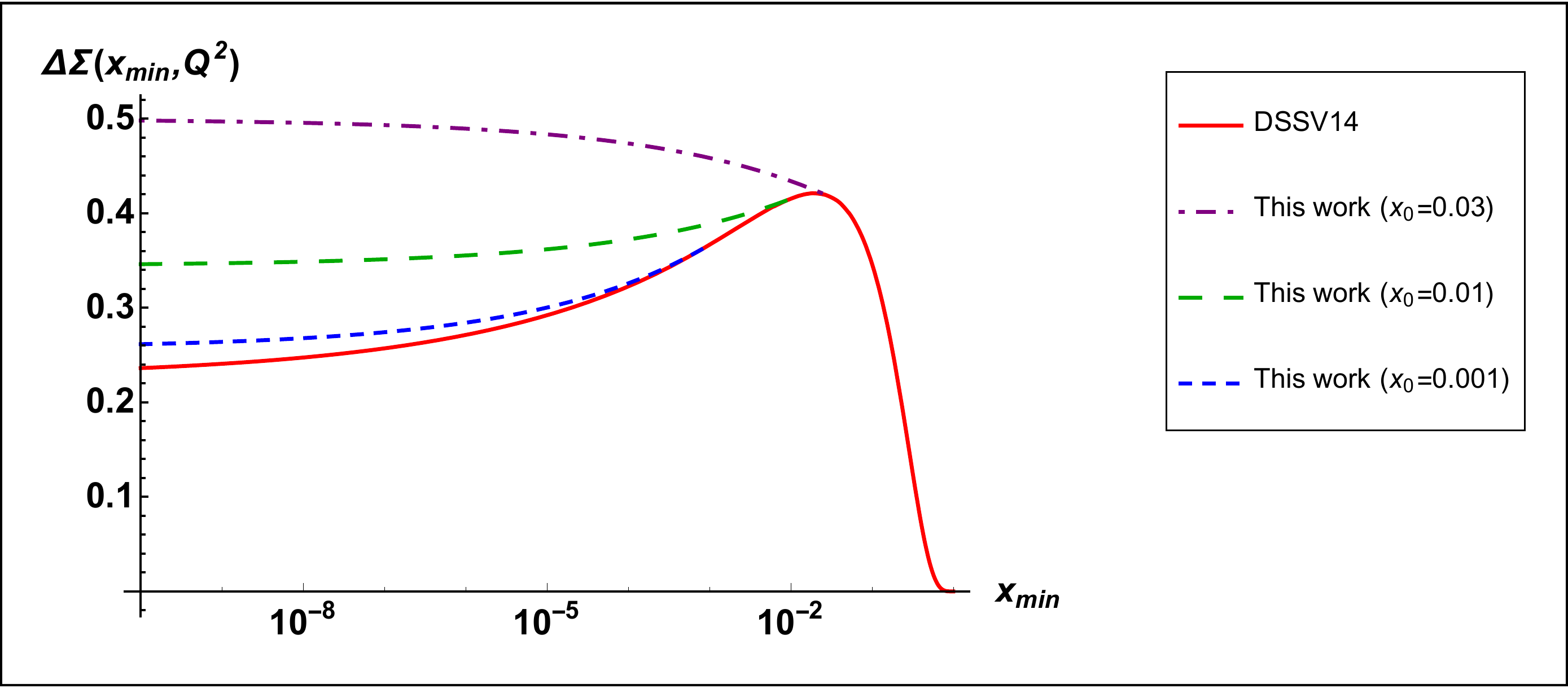} 
\caption{Plot of $\Delta\Sigma^{[x_{min}]}\!(Q^2)$ vs.~$x_{min}$ at
  $Q^2=10\,{\rm GeV}^2$.  The solid curve is from
  DSSV14~\cite{deFlorian:2014yva}.  The dot-dashed, long-dashed, and
  short-dashed curves are from various small-$x$ modifications of
  $\Delta\Sigma(x,Q^2)$ at $x_0=0.03,\,0.01,\,0.001$, respectively,
  using our helicity intercept (see the text for
  details).} \label{f:run_spin}
\end{figure}

We now will attempt to quantify how the small-$x$ behavior of
$\Delta\Sigma(x,Q^2)$ derived here affects the integral in
Eq.~(\ref{eq:net_spin}).  We take a simple approach and leave a more
rigorous phenomenological study for future work.  First, we attach a
curve $\Delta\tilde{\Sigma}(x,Q^2) = N\,x^{-\alpha_h}$ (with
$\alpha_h$ given in (\ref{e:alphah})) to the DSSV14 result for
$\Delta\Sigma(x,Q^2)$ at a particular small-$x$ point $x_0$.  Next, we
fix the normalization $N$ by requiring $\Delta\tilde{\Sigma}(x_0,Q^2)
= \Delta\Sigma(x_0,Q^2)$.  Finally, we calculate the truncated
integral \eqref{e:run_int} of the modified quark helicity PDF
\begin{align} 
  \label{e:run_int_mod}
  \Delta\Sigma_{mod}(x,Q^2) \equiv & \ \theta(x-x_{0}) \,
  \Delta\Sigma(x,Q^2) \notag \\ & + \theta(x_0-x) \,
  \Delta\tilde{\Sigma}(x,Q^2)
\end{align}
for different $x_0$ values. The results are shown in
Fig.~\ref{f:run_spin} for $Q^2 = 10\,{\rm GeV^2}$ and $\alpha_s\approx
0.25$, in which case $\alpha_h \approx 0.80$.

We see that the small-$x$ evolution of $\Delta\Sigma(x,Q^2)$ could
offer a moderate to significant enhancement to the quark spin,
depending on where in $x$ the effects set in and on the
parameterization of the helicity PDFs at higher $x$.  Thus, it will be
important to incorporate the results of this work, and more generally
the small-$x$ helicity evolution equations discussed here, into future
extractions of helicity PDFs.


\section{Summary}

In \cite{Kovchegov:2015pbl,Kovchegov:2016zex} we have derived
small-$x$ evolution equations for the polarized dipole amplitude. The
equations close in the large-$N_c$ and large-$N_c \, \& \, N_f$
limits. The large-$N_c$ equations are presented above. The solution
of these equations provides theoretical input on the perturbative
value of the small-$x$ intercept for the quark helicity TMD and PDF,
and for the $g_1$ structure function.

We have also numerically solved the small-$x$ helicity evolution
equations \eqref{evol88} of Ref.~\cite{Kovchegov:2015pbl} in the
large-$N_c$ limit.  We found an intercept of $\alpha_h=2.31\sqrt{\as
  N_c/2\pi}$, which, from Eq.~(\ref{e:DeltaSigma}), is a direct input
from theory on the behavior of $\Delta\Sigma(x,Q^2)$ at small $x$.
Although a more rigorous phenomenological study is needed, we
demonstrated in a simple approach that such an intercept could offer a
moderate to significant enhancement of the quark contribution to the
proton spin.  Therefore, it appears imperative to include the effects
of the small-$x$ helicity evolution discussed here in future fits of
helicity PDFs, especially those to be obtained at an Electron-Ion
Collider.


\begin{acknowledgments}

  This material is based upon work supported by the U.S. Department of
  Energy, Office of Science, Office of Nuclear Physics under Award
  Number DE-SC0004286 (YK) and within the framework of the TMD Topical
  Collaboration (DP) and DOE Contract No.~DE-SC0012704 (MS). MS
  received additional support from an EIC program development fund
  from BNL and from the U.S. Department of Energy, Office of Science
  under the DOE Early Career Program.

\end{acknowledgments}




\begin{thebibliography}{17}%
\makeatletter
\providecommand \@ifxundefined [1]{%
 \@ifx{#1\undefined}
}%
\providecommand \@ifnum [1]{%
 \ifnum #1\expandafter \@firstoftwo
 \else \expandafter \@secondoftwo
 \fi
}%
\providecommand \@ifx [1]{%
 \ifx #1\expandafter \@firstoftwo
 \else \expandafter \@secondoftwo
 \fi
}%
\providecommand \natexlab [1]{#1}%
\providecommand \enquote  [1]{``#1''}%
\providecommand \bibnamefont  [1]{#1}%
\providecommand \bibfnamefont [1]{#1}%
\providecommand \citenamefont [1]{#1}%
\providecommand \href@noop [0]{\@secondoftwo}%
\providecommand \href [0]{\begingroup \@sanitize@url \@href}%
\providecommand \@href[1]{\@@startlink{#1}\@@href}%
\providecommand \@@href[1]{\endgroup#1\@@endlink}%
\providecommand \@sanitize@url [0]{\catcode `\\12\catcode `\$12\catcode
  `\&12\catcode `\#12\catcode `\^12\catcode `\_12\catcode `\%12\relax}%
\providecommand \@@startlink[1]{}%
\providecommand \@@endlink[0]{}%
\providecommand \url  [0]{\begingroup\@sanitize@url \@url }%
\providecommand \@url [1]{\endgroup\@href {#1}{\urlprefix }}%
\providecommand \urlprefix  [0]{URL }%
\providecommand \Eprint [0]{\href }%
\providecommand \doibase [0]{http://dx.doi.org/}%
\providecommand \selectlanguage [0]{\@gobble}%
\providecommand \bibinfo  [0]{\@secondoftwo}%
\providecommand \bibfield  [0]{\@secondoftwo}%
\providecommand \translation [1]{[#1]}%
\providecommand \BibitemOpen [0]{}%
\providecommand \bibitemStop [0]{}%
\providecommand \bibitemNoStop [0]{.\EOS\space}%
\providecommand \EOS [0]{\spacefactor3000\relax}%
\providecommand \BibitemShut  [1]{\csname bibitem#1\endcsname}%
\let\auto@bib@innerbib\@empty
\bibitem [{\citenamefont {Kovchegov}\ \emph
  {et~al.}(2016{\natexlab{a}})\citenamefont {Kovchegov}, \citenamefont
  {Pitonyak},\ and\ \citenamefont {Sievert}}]{Kovchegov:2015pbl}%
  \BibitemOpen
  \bibfield  {author} {\bibinfo {author} {\bibfnamefont {Y.~V.}\ \bibnamefont
  {Kovchegov}}, \bibinfo {author} {\bibfnamefont {D.}~\bibnamefont {Pitonyak}},
  \ and\ \bibinfo {author} {\bibfnamefont {M.~D.}\ \bibnamefont {Sievert}},\
  }\href {\doibase 10.1007/JHEP01(2016)072} {\bibfield  {journal} {\bibinfo
  {journal} {JHEP}\ }\textbf {\bibinfo {volume} {01}},\ \bibinfo {pages} {072}
  (\bibinfo {year} {2016}{\natexlab{a}})},\ \Eprint
  {http://arxiv.org/abs/1511.06737} {arXiv:1511.06737 [hep-ph]} \BibitemShut
  {NoStop}%
\bibitem [{\citenamefont {Kovchegov}\ \emph
  {et~al.}(2016{\natexlab{b}})\citenamefont {Kovchegov}, \citenamefont
  {Pitonyak},\ and\ \citenamefont {Sievert}}]{Kovchegov:2016weo}%
  \BibitemOpen
  \bibfield  {author} {\bibinfo {author} {\bibfnamefont {Y.~V.}\ \bibnamefont
  {Kovchegov}}, \bibinfo {author} {\bibfnamefont {D.}~\bibnamefont {Pitonyak}},
  \ and\ \bibinfo {author} {\bibfnamefont {M.~D.}\ \bibnamefont {Sievert}},\
  }\href@noop {} {\  (\bibinfo {year} {2016}{\natexlab{b}})},\ \Eprint
  {http://arxiv.org/abs/1610.06188} {arXiv:1610.06188 [hep-ph]} \BibitemShut
  {NoStop}%
\bibitem [{\citenamefont {Kovchegov}\ \emph
  {et~al.}(2016{\natexlab{c}})\citenamefont {Kovchegov}, \citenamefont
  {Pitonyak},\ and\ \citenamefont {Sievert}}]{Kovchegov:2016zex}%
  \BibitemOpen
  \bibfield  {author} {\bibinfo {author} {\bibfnamefont {Y.~V.}\ \bibnamefont
  {Kovchegov}}, \bibinfo {author} {\bibfnamefont {D.}~\bibnamefont {Pitonyak}},
  \ and\ \bibinfo {author} {\bibfnamefont {M.~D.}\ \bibnamefont {Sievert}},\
  }\href@noop {} {\  (\bibinfo {year} {2016}{\natexlab{c}})},\ \Eprint
  {http://arxiv.org/abs/1610.06197} {arXiv:1610.06197 [hep-ph]} \BibitemShut
  {NoStop}%
\bibitem [{\citenamefont {Kovchegov}\ and\ \citenamefont
  {Sievert}(2016)}]{Kovchegov:2015zha}%
  \BibitemOpen
  \bibfield  {author} {\bibinfo {author} {\bibfnamefont {Y.~V.}\ \bibnamefont
  {Kovchegov}}\ and\ \bibinfo {author} {\bibfnamefont {M.~D.}\ \bibnamefont
  {Sievert}},\ }\href {\doibase 10.1016/j.nuclphysb.2015.12.008} {\bibfield
  {journal} {\bibinfo  {journal} {Nucl. Phys.}\ }\textbf {\bibinfo {volume}
  {B903}},\ \bibinfo {pages} {164} (\bibinfo {year} {2016})},\ \Eprint
  {http://arxiv.org/abs/1505.01176} {arXiv:1505.01176 [hep-ph]} \BibitemShut
  {NoStop}%
\bibitem [{\citenamefont {Itakura}\ \emph {et~al.}(2004)\citenamefont
  {Itakura}, \citenamefont {Kovchegov}, \citenamefont {McLerran},\ and\
  \citenamefont {Teaney}}]{Itakura:2003jp}%
  \BibitemOpen
  \bibfield  {author} {\bibinfo {author} {\bibfnamefont {K.}~\bibnamefont
  {Itakura}}, \bibinfo {author} {\bibfnamefont {Y.~V.}\ \bibnamefont
  {Kovchegov}}, \bibinfo {author} {\bibfnamefont {L.}~\bibnamefont {McLerran}},
  \ and\ \bibinfo {author} {\bibfnamefont {D.}~\bibnamefont {Teaney}},\ }\href
  {\doibase 10.1016/j.nuclphysa.2003.10.016} {\bibfield  {journal} {\bibinfo
  {journal} {Nucl. Phys.}\ }\textbf {\bibinfo {volume} {A730}},\ \bibinfo
  {pages} {160} (\bibinfo {year} {2004})},\ \Eprint
  {http://arxiv.org/abs/hep-ph/0305332} {arXiv:hep-ph/0305332} \BibitemShut
  {NoStop}%
\bibitem [{\citenamefont {Bartels}\ \emph {et~al.}(1996)\citenamefont
  {Bartels}, \citenamefont {Ermolaev},\ and\ \citenamefont
  {Ryskin}}]{Bartels:1996wc}%
  \BibitemOpen
  \bibfield  {author} {\bibinfo {author} {\bibfnamefont {J.}~\bibnamefont
  {Bartels}}, \bibinfo {author} {\bibfnamefont {B.}~\bibnamefont {Ermolaev}}, \
  and\ \bibinfo {author} {\bibfnamefont {M.}~\bibnamefont {Ryskin}},\ }\href
  {\doibase 10.1007/s002880050285} {\bibfield  {journal} {\bibinfo  {journal}
  {Z.Phys.}\ }\textbf {\bibinfo {volume} {C72}},\ \bibinfo {pages} {627}
  (\bibinfo {year} {1996})},\ \Eprint {http://arxiv.org/abs/hep-ph/9603204}
  {arXiv:hep-ph/9603204 [hep-ph]} \BibitemShut {NoStop}%
\bibitem [{\citenamefont {Kovchegov}\ and\ \citenamefont
  {Levin}(2012)}]{KovchegovLevin}%
  \BibitemOpen
  \bibfield  {author} {\bibinfo {author} {\bibfnamefont {Y.~V.}\ \bibnamefont
  {Kovchegov}}\ and\ \bibinfo {author} {\bibfnamefont {E.}~\bibnamefont
  {Levin}},\ }\href@noop {} {\emph {\bibinfo {title} {Quantum Chromodynamics at
  High Energy}}}\ (\bibinfo  {publisher} {Cambridge University Press},\
  \bibinfo {year} {2012})\BibitemShut {NoStop}%
\bibitem [{\citenamefont {de~Florian}\ \emph {et~al.}(2009)\citenamefont
  {de~Florian}, \citenamefont {Sassot}, \citenamefont {Stratmann},\ and\
  \citenamefont {Vogelsang}}]{deFlorian:2009vb}%
  \BibitemOpen
  \bibfield  {author} {\bibinfo {author} {\bibfnamefont {D.}~\bibnamefont
  {de~Florian}}, \bibinfo {author} {\bibfnamefont {R.}~\bibnamefont {Sassot}},
  \bibinfo {author} {\bibfnamefont {M.}~\bibnamefont {Stratmann}}, \ and\
  \bibinfo {author} {\bibfnamefont {W.}~\bibnamefont {Vogelsang}},\ }\href
  {\doibase 10.1103/PhysRevD.80.034030} {\bibfield  {journal} {\bibinfo
  {journal} {Phys. Rev.}\ }\textbf {\bibinfo {volume} {D80}},\ \bibinfo {pages}
  {034030} (\bibinfo {year} {2009})},\ \Eprint {http://arxiv.org/abs/0904.3821}
  {arXiv:0904.3821 [hep-ph]} \BibitemShut {NoStop}%
\bibitem [{\citenamefont {de~Florian}\ \emph {et~al.}(2014)\citenamefont
  {de~Florian}, \citenamefont {Sassot}, \citenamefont {Stratmann},\ and\
  \citenamefont {Vogelsang}}]{deFlorian:2014yva}%
  \BibitemOpen
  \bibfield  {author} {\bibinfo {author} {\bibfnamefont {D.}~\bibnamefont
  {de~Florian}}, \bibinfo {author} {\bibfnamefont {R.}~\bibnamefont {Sassot}},
  \bibinfo {author} {\bibfnamefont {M.}~\bibnamefont {Stratmann}}, \ and\
  \bibinfo {author} {\bibfnamefont {W.}~\bibnamefont {Vogelsang}},\ }\href
  {\doibase 10.1103/PhysRevLett.113.012001} {\bibfield  {journal} {\bibinfo
  {journal} {Phys. Rev. Lett.}\ }\textbf {\bibinfo {volume} {113}},\ \bibinfo
  {pages} {012001} (\bibinfo {year} {2014})},\ \Eprint
  {http://arxiv.org/abs/1404.4293} {arXiv:1404.4293 [hep-ph]} \BibitemShut
  {NoStop}%
\bibitem [{\citenamefont {Jimenez-Delgado}\ \emph {et~al.}(2014)\citenamefont
  {Jimenez-Delgado}, \citenamefont {Accardi},\ and\ \citenamefont
  {Melnitchouk}}]{Jimenez-Delgado:2013boa}%
  \BibitemOpen
  \bibfield  {author} {\bibinfo {author} {\bibfnamefont {P.}~\bibnamefont
  {Jimenez-Delgado}}, \bibinfo {author} {\bibfnamefont {A.}~\bibnamefont
  {Accardi}}, \ and\ \bibinfo {author} {\bibfnamefont {W.}~\bibnamefont
  {Melnitchouk}},\ }\href {\doibase 10.1103/PhysRevD.89.034025} {\bibfield
  {journal} {\bibinfo  {journal} {Phys. Rev.}\ }\textbf {\bibinfo {volume}
  {D89}},\ \bibinfo {pages} {034025} (\bibinfo {year} {2014})},\ \Eprint
  {http://arxiv.org/abs/1310.3734} {arXiv:1310.3734 [hep-ph]} \BibitemShut
  {NoStop}%
\bibitem [{\citenamefont {Sato}\ \emph {et~al.}(2016)\citenamefont {Sato},
  \citenamefont {Melnitchouk}, \citenamefont {Kuhn}, \citenamefont {Ethier},\
  and\ \citenamefont {Accardi}}]{Sato:2016tuz}%
  \BibitemOpen
  \bibfield  {author} {\bibinfo {author} {\bibfnamefont {N.}~\bibnamefont
  {Sato}}, \bibinfo {author} {\bibfnamefont {W.}~\bibnamefont {Melnitchouk}},
  \bibinfo {author} {\bibfnamefont {S.~E.}\ \bibnamefont {Kuhn}}, \bibinfo
  {author} {\bibfnamefont {J.~J.}\ \bibnamefont {Ethier}}, \ and\ \bibinfo
  {author} {\bibfnamefont {A.}~\bibnamefont {Accardi}} (\bibinfo
  {collaboration} {Jefferson Lab Angular Momentum}),\ }\href {\doibase
  10.1103/PhysRevD.93.074005} {\bibfield  {journal} {\bibinfo  {journal} {Phys.
  Rev.}\ }\textbf {\bibinfo {volume} {D93}},\ \bibinfo {pages} {074005}
  (\bibinfo {year} {2016})},\ \Eprint {http://arxiv.org/abs/1601.07782}
  {arXiv:1601.07782 [hep-ph]} \BibitemShut {NoStop}%
\bibitem [{\citenamefont {Leader}\ \emph {et~al.}(2006)\citenamefont {Leader},
  \citenamefont {Sidorov},\ and\ \citenamefont {Stamenov}}]{Leader:2005ci}%
  \BibitemOpen
  \bibfield  {author} {\bibinfo {author} {\bibfnamefont {E.}~\bibnamefont
  {Leader}}, \bibinfo {author} {\bibfnamefont {A.~V.}\ \bibnamefont {Sidorov}},
  \ and\ \bibinfo {author} {\bibfnamefont {D.~B.}\ \bibnamefont {Stamenov}},\
  }\href {\doibase 10.1103/PhysRevD.73.034023} {\bibfield  {journal} {\bibinfo
  {journal} {Phys. Rev.}\ }\textbf {\bibinfo {volume} {D73}},\ \bibinfo {pages}
  {034023} (\bibinfo {year} {2006})},\ \Eprint
  {http://arxiv.org/abs/hep-ph/0512114} {arXiv:hep-ph/0512114 [hep-ph]}
  \BibitemShut {NoStop}%
\bibitem [{\citenamefont {Leader}\ \emph {et~al.}(2010)\citenamefont {Leader},
  \citenamefont {Sidorov},\ and\ \citenamefont {Stamenov}}]{Leader:2010rb}%
  \BibitemOpen
  \bibfield  {author} {\bibinfo {author} {\bibfnamefont {E.}~\bibnamefont
  {Leader}}, \bibinfo {author} {\bibfnamefont {A.~V.}\ \bibnamefont {Sidorov}},
  \ and\ \bibinfo {author} {\bibfnamefont {D.~B.}\ \bibnamefont {Stamenov}},\
  }\href {\doibase 10.1103/PhysRevD.82.114018} {\bibfield  {journal} {\bibinfo
  {journal} {Phys. Rev.}\ }\textbf {\bibinfo {volume} {D82}},\ \bibinfo {pages}
  {114018} (\bibinfo {year} {2010})},\ \Eprint {http://arxiv.org/abs/1010.0574}
  {arXiv:1010.0574 [hep-ph]} \BibitemShut {NoStop}%
\bibitem [{\citenamefont {Leader}\ \emph {et~al.}(2015)\citenamefont {Leader},
  \citenamefont {Sidorov},\ and\ \citenamefont {Stamenov}}]{Leader:2014uua}%
  \BibitemOpen
  \bibfield  {author} {\bibinfo {author} {\bibfnamefont {E.}~\bibnamefont
  {Leader}}, \bibinfo {author} {\bibfnamefont {A.~V.}\ \bibnamefont {Sidorov}},
  \ and\ \bibinfo {author} {\bibfnamefont {D.~B.}\ \bibnamefont {Stamenov}},\
  }\href {\doibase 10.1103/PhysRevD.91.054017} {\bibfield  {journal} {\bibinfo
  {journal} {Phys. Rev.}\ }\textbf {\bibinfo {volume} {D91}},\ \bibinfo {pages}
  {054017} (\bibinfo {year} {2015})},\ \Eprint {http://arxiv.org/abs/1410.1657}
  {arXiv:1410.1657 [hep-ph]} \BibitemShut {NoStop}%
\bibitem [{\citenamefont {Ball}\ \emph {et~al.}(2013)\citenamefont {Ball},
  \citenamefont {Forte}, \citenamefont {Guffanti}, \citenamefont {Nocera},
  \citenamefont {Ridolfi},\ and\ \citenamefont {Rojo}}]{Ball:2013lla}%
  \BibitemOpen
  \bibfield  {author} {\bibinfo {author} {\bibfnamefont {R.~D.}\ \bibnamefont
  {Ball}}, \bibinfo {author} {\bibfnamefont {S.}~\bibnamefont {Forte}},
  \bibinfo {author} {\bibfnamefont {A.}~\bibnamefont {Guffanti}}, \bibinfo
  {author} {\bibfnamefont {E.~R.}\ \bibnamefont {Nocera}}, \bibinfo {author}
  {\bibfnamefont {G.}~\bibnamefont {Ridolfi}}, \ and\ \bibinfo {author}
  {\bibfnamefont {J.}~\bibnamefont {Rojo}} (\bibinfo {collaboration} {NNPDF}),\
  }\href {\doibase 10.1016/j.nuclphysb.2013.05.007} {\bibfield  {journal}
  {\bibinfo  {journal} {Nucl. Phys.}\ }\textbf {\bibinfo {volume} {B874}},\
  \bibinfo {pages} {36} (\bibinfo {year} {2013})},\ \Eprint
  {http://arxiv.org/abs/1303.7236} {arXiv:1303.7236 [hep-ph]} \BibitemShut
  {NoStop}%
\bibitem [{\citenamefont {Nocera}\ \emph {et~al.}(2014)\citenamefont {Nocera},
  \citenamefont {Ball}, \citenamefont {Forte}, \citenamefont {Ridolfi},\ and\
  \citenamefont {Rojo}}]{Nocera:2014gqa}%
  \BibitemOpen
  \bibfield  {author} {\bibinfo {author} {\bibfnamefont {E.~R.}\ \bibnamefont
  {Nocera}}, \bibinfo {author} {\bibfnamefont {R.~D.}\ \bibnamefont {Ball}},
  \bibinfo {author} {\bibfnamefont {S.}~\bibnamefont {Forte}}, \bibinfo
  {author} {\bibfnamefont {G.}~\bibnamefont {Ridolfi}}, \ and\ \bibinfo
  {author} {\bibfnamefont {J.}~\bibnamefont {Rojo}} (\bibinfo {collaboration}
  {NNPDF}),\ }\href {\doibase 10.1016/j.nuclphysb.2014.08.008} {\bibfield
  {journal} {\bibinfo  {journal} {Nucl. Phys.}\ }\textbf {\bibinfo {volume}
  {B887}},\ \bibinfo {pages} {276} (\bibinfo {year} {2014})},\ \Eprint
  {http://arxiv.org/abs/1406.5539} {arXiv:1406.5539 [hep-ph]} \BibitemShut
  {NoStop}%
\bibitem [{\citenamefont {Aschenauer}\ \emph {et~al.}(2015)\citenamefont
  {Aschenauer}, \citenamefont {Sassot},\ and\ \citenamefont
  {Stratmann}}]{Aschenauer:2015ata}%
  \BibitemOpen
  \bibfield  {author} {\bibinfo {author} {\bibfnamefont {E.~C.}\ \bibnamefont
  {Aschenauer}}, \bibinfo {author} {\bibfnamefont {R.}~\bibnamefont {Sassot}},
  \ and\ \bibinfo {author} {\bibfnamefont {M.}~\bibnamefont {Stratmann}},\
  }\href {\doibase 10.1103/PhysRevD.92.094030} {\bibfield  {journal} {\bibinfo
  {journal} {Phys. Rev.}\ }\textbf {\bibinfo {volume} {D92}},\ \bibinfo {pages}
  {094030} (\bibinfo {year} {2015})},\ \Eprint
  {http://arxiv.org/abs/1509.06489} {arXiv:1509.06489 [hep-ph]} \BibitemShut
  {NoStop}%
\end{thebibliography}

%

\end{document}